# Synthesis, crystal structure, and properties of novel perovskite oxychalcogenides, $Ca_2CuFeO_3Ch$ (Ch = S, Se)

D.O. Charkin[1*], A.V. Sadakov[2], O.E. Omel'yanovskii[2], and S.M. Kazakov[1]

[1] *Department of Chemistry, M.V. Lomonosov Moscow State University, 119991, Moscow, Russia*

[2] *P.N. Lebedev Physical Institute, RAS, Moscow, 119991, Russia*

\* - corresponding author, e-mail: charkin@inorg.chem.msu.ru

**Abstract**

Two new perovskite oxychalcogenides, $Ca_2CuFeO_3S$ and $Ca_2CuFeO_3Se$, have been synthesized in evacuated quartz tubes. They crystallize in *P4/nmm* space group with lattice parameters *a* = 3.8271(1), *c* = 14.9485(2) Å and *a* = 3.8605(1), *c* = 15.3030(2) Å for $Ca_2CuFeO_3S$ and $Ca_2CuFeO_3Se$, respectively. They appear to be the first layered chalcogenide perovskites involving calcium and are structural analogs of the corresponding Sr and Ba compounds. The new compounds exhibit semiconducting properties with energy gap decreasing from the oxysulfide to the oxyselenide. Possibility of introducing $Ca^{2+}$ into structures of known layered oxychalcogenides and oxypnictides is discussed.

**Keywords:** A. Layered compounds; B. Chemical synthesis, C. X-ray diffraction; D. Crystal structure

## 1. Introduction

The recent discovery of high-temperature superconductivity in an oxypnictide La(O,F)FeAs with a relatively simple LaOAgS structure [1] has triggered a burst of interest in search of novel representatives of known and novel structure types bearing [$Fe_2As_2$] and other *anti*-fluorite layers [2–4]. Most of this work has hitherto been done by analogy with the structures and compositions of oxychalcogenides, a family intimately related, both structurally and chemically, to the target pnictides. To the moment, the oxychalcogenides have been studied much more thoroughly as promising transparent semiconductors and ionic conductors (see the review [2] and references therein). It should be noted, however, that most structures of the oxychalcogenides and oxypnictides have been in their turn obtained by filling empty tetrahedral voids in the structures of layered oxyhalides, especially perovskite-derived, as has been pointed out earlier [2, 5, 6]. For instance, the structures of the most promising superconductors, i.e. Ln(O,F)FeAs [1], $Sr_2VFeO_3As$ [3], and $Sr_3Sc_2Fe_2O_5As_2$ [4], are, in fact, the "filled-up" versions of the LnOCl (PbFCl) [6], $Ca_2FeO_3Cl$ [7], and $Sr_3Fe_2O_5Cl_2$ [7, 8] (Fig. 1). Thus oxyhalides may prove to be a good source of new charge reservoirs, both for new oxychalcogenides and oxypnictides. Even a cursory survey of the oxide halide family shows that whereas $Ca^{2+}$ is a typical constituent of perovskite blocks therein (mostly together with $Cu^{2+}$ or $Fe^{3+}$ at the *B*-site [8–10]), this alkaline earth cation had not yet been used to contribute to the chalcogenides. Two series of calcium-containing superconducting oxypnictides, $Ca_{n+1}(Sc,Ti)_nO_{3n-1}Fe_2As_2$ [11] and $Ca_{n+1}(Mg,Ti)_nOFe_2As_2$ [12, 13], have been prepared recently, but most promising structure types, $Sr_2VFeO_3As$ ($Sr_2CuGaO_3S$) and $Sr_3Sc_2Fe_2O_5As_2$ ($Sr_3Fe_2Cu_2O_5S_2$), remain yet unaddressed. The calcium-bearing oxyhalides were as yet obtained only among compounds of Cu and Fe. As both $Fe^{3+}$ and $Cu^{2+}$ are redox-incompatible with the pnictide anions, the search for new calcium-bearing compounds has been conducted among oxychalcogenides. In the present paper, we report preparation, crystal structures and physical properties of the novel calcium compounds, $Ca_2CuFeO_3S$ and $Ca_2FeO_3Se$, with the $Sr_2CuGaO_3S$-type [14] structure. They are structural analogs of the corresponding Sr-containing compound $Sr_2CuFeO_3S$, first reported by Zhu and Hor [15].

## 2. Experimental

### 2.1. Synthesis

The starting chemicals were CaO (calcined at 1100°C for 48hrs), CuO (obtained by thermal decomposition of $Cu(NO_3)_2 \cdot 3H_2O$ at 400°C for 2 hrs), Fe (freshly reduced by hydrogen at 400°C for 1 hr), sulfur (melted *in vacuo* for 15 min to remove traces of water), selenium and tellurium (which were used as purchased). All chemicals were of analytically or extra pure grade. The studies were initially aimed at both possible stoichiometries, i.e. $Ca_2CuFeO_3Ch$ and $Ca_3Cu_2Fe_2O_5Ch_2$. Thus, mixtures of CaO, CuO, Fe, and chalcogen in 2:1:1:1 or 3:2:2:2 ratios were thoroughly ground, pressed into pellets (at ca. 8 ton/cm$^2$ for 60 sec), sealed in evacuated quartz tubes (at the residual pressure of ca. 0.05 Torr), and annealed twice at 675°C for 36 hrs for the $Ca_2CuFeO_3Ch$ compositions, or 750°C for 48 hrs for the $Ca_3Cu_2Fe_2O_5Ch_2$-aimed samples, with one intermediate re-grinding and re-pelletizing. The sulfide samples were pre-annealed at 400°C for one day to ensure complete reaction of sulfur.

### 2.2. Characterization

**Powder X-ray diffraction** data were collected on a STOE STADI/P diffractometer (Cu-$K_{\alpha1}$ radiation, Ge-111 monochromator, transmission geometry) between 2θ=5-110° at steps of 0.02° and counting time of 10 s. Ritveld refinements of both samples were performed with the TOPAS package [16] using pseudo-Voigt peak shape function. The structure of the $Sr_2CuGaO_3S$ [14] was used for the starting coordinate values. Preferred orientation along [001] was corrected using a March-Dollase function.

**Magnetic measurements** were performed using a Quantum Design PPMS ac-susceptometer. In this standard method an alternating magnetic field is applied to the sample via copper drive coil, and a detection coil set (two counterwound copper coils connected in series) inductively responds to the combined sample moment and excitation field. The sample was placed inside one of the detection coils. Amplitude and frequency of the applied ac-field were 1 Oe and 137 Hz, respectively. The samples were cooled down in zero dc-field.

**Resistivity measurements** were performed using the standard 4-probe technique with a help of Stanford Research SR830 lock-in amplifier and SR560 preamplifier. Due to high sample resistance, we applied ac-currents as small as 10 nA. For low temperature measurements, a Dewar insert with CuFeCu thermocouple and a Lakeshore 340 temperature controller were used. For high temperature measurements, we used a high temperature insert, an oven, and a Cu-Constantan-Cu thermocouple. The temperature resolution of these thermocouples was ~10μV/K and ~20 μV/K, respectively.

### 3. Results and Discussion

#### 3.1. New compounds and their properties

The X-ray diffraction data indicated formation of two target products, $Ca_2CuFeO_3S$ and $Ca_2CuFeO_3Se$, brown in color. These phases are isostructural to $Sr_2CuGaO_3S$ with two square-pyramidal oxide layers alternating with $Cu_2Ch_2$ (Ch=S, Se) block. A single phase was obtained for $Ca_2CuFeO_3S$. Some minor impurity peaks originating from $CuFeSe_2$ [17] were found in the X-ray diffraction pattern of $Ca_2CuFeO_3Se$ and this phase was also included in the refinement. Details of the Rietveld refinements of both samples are listed in Table 1. Refined atomic positions are given in Table 2. The resulting bond distances are collected in Table 3 together with the data for related and isostructural compounds for comparison. Final Rietveld refinement plots are presented in Fig. 2.

In the $Ca_3Cu_2Fe_2O_5Ch_2$-aimed samples, only $Ca_2Fe_2O_5$, CaCh (Ch=S, Se), and $Cu_2Ch$ (Ch=S, Se) were observed after synthesis. In the case of samples containing tellurium no target compounds were observed. Attempts were also made to prepare analogous $Ca_2CuMO_3Ch$ (M = Cr, Mn) and $Ca_2AgFeO_3Ch$ compounds (under similar conditions), but yet also without success.

Temperature dependences of magnetic susceptibility of the synthesized samples are presented on Fig. 3(a) and 3(b). The two panels show two components (in-phase M′ and out-of phase M″) of the ac-signal. For both samples, the in-phase component M′ exhibits extrema and inflection as a function of temperature, whereas the out-of-phase component, which shows energy losses in the sample, exhibits several peaks. Obviously, there is an anti-ferromagnetic transition with a Néel temperature $T_N$ of ca. 150 K, also there are at least two transitions at lower temperatures (around 50 K and 30 K). Their nature is yet to be clarified by additional studies. If we compare our results to those for a similar compound $Sr_3Cu_2Fe_2O_5S_2$ [18], we find a more complex behavior in our case.

Temperature dependences of the resistivity for the two samples are shown on Figs. 4(a) and 4(b). The $Ca_2CuFeO_3S$ sample with an extremely high resistance was measured only in the temperature range from 260 K to 340 K. Its $\rho(T)$ dependence shows an approximately temperature activated type behavior, with activation energy about 0.17eV. The deviation from the logarithmic law, obviously, indicates a change in type of conductivity The $Ca_2CuFeO_3Se$ sample had a lower resistance and its $\rho(T)$ dependence was measured in the wider range 14-300 K. As seen from inset of Fig. 5(b), its R(T) dependence may be approximated with $R \propto \exp[(T_0/T)^{1/4}]$, which is typical for variable-range hopping transport.

### 3.2. Crystal chemistry

The two new compounds, $Ca_2CuFeO_3S$ and $Ca_2CuFeO_3Se$, belong to the "42262" ($Sr_2CuGaO_3S \equiv Sr_4Cu_2Ga_2O_6S_2$) structure type. Upon comparison of structural data for $Ca_2CuFeO_3S$ and $Ca_2FeO_3Cl$, one can see that the *a* cell parameters do not change significantly. It is also the case when passing from $Sr_2FeO_3Cl$ to $Sr_2CuFeO_3S$. This indicates that the $[Ae_2FeO_3]^+$ (Ae = Ca, Sr) block is relatively rigid, and the *a* cell parameters are more or less transferable within the structures involving them. This is not so much true for the Ca–O distances which change essentially from $Ca_2FeO_3Cl$ to $Ca_2CuFeO_3S$. The bond length distribution in $Ca_2FeO_3Cl$, however, is broader as compared to both $Sr_2CuFeO_3S$ and $Ca_2CuFeO_3Ch$. The reason may appear from differences in the strength of Ae–X and Ae–Ch bonds; unfortunately, there are no other data which permit direct comparison of $Ae_2MO_3X$ and $Ae_2CuMO_3Ch$ for the same couple of Ae and M. There is a small elongation of all Ca–O distances when passing from the sulfide to the corresponding selenide, as well as elongation of Cu–S distance when passing from $Ca_2CuFeO_3S$ to $Sr_2FeCuO_3S$. This can be explained by increase of the *a* cell parameter when passing from Ca to Sr, or from S to Se.

A close look at Table 2 reveals a characteristic feature for the compounds bearing $[Cu_2X_2]$ *anti*-fluorite slabs (X = chalcogen or pnicogen), the unusually large thermal parameter for the $Cu^+$ cations which suggests slight non-stoichiometry at this site. This issue has been addressed for a long time, and recently proven to occur due to reversible copper extraction resulting from oxidation by moist air [2]. The high thermal parameters of Cu in the $Ca_2CuFeO_3Ch$ structures most likely illustrate the same phenomenon.

Our results confirm that the chemistry of layered perovskite oxide chalcogenides and pnictides is indeed not restricted to compounds of Sr and Ba, and calcium appears to be, at least within some structure types, just their smaller analog akin to the oxide halide family [8]. Considering the tendency of increasing the $T_C$ of superconducting iron pnictides with decrease of cell dimensions in the *ab* plane (*e.g.* in the LnOFeAs family [19]), existence of calcium compounds is likely to provide another tool for chemical compression of the superconducting layers, al least in layered perovskites. With calcium, we succeeded as yet in preparation of only $Fe^{III}$ compounds, and this oxidation state is not directly transferable to pnictides due to redox incompatibility of $Fe^{III}$ and $Pn^{-III}$ [20]. Further studies of calcium compounds involving cations stable to reduction, e.g. $Ga^{III}$, are evidently necessary before we see whether isostructural Ca oxide pnictides can exist.

Unfortunately, the direct relationships between composition and structures are yet unknown as evidenced by the fact that the suggested $Ca_3Ch_2Fe_2O_5Ch_2$ compounds were not obtained. Though the differences in the *a* cell parameters between $Ca_2FeO_3Cl$ and $Ca_3Fe_2O_5Cl_2$ are marginal, the double anionic layers in the former compound may be stuffed with $Cu^+$ while in the latter, not as yet. Unless proper synthetic conditions have not been found, there seems to be a rather delicate interplay between the "42262" ($Sr_2CuGaO_3S$) and "32252" ($Sr_3Cu_2Fe_2O_5S_2$) structures both in the oxide chalcogenide and oxide pnictide families, the possibility of formation depending on the nature of both perovskite and *anti*-fluorite layers. Among compounds of Sr and Ba, for instance, the "42622" structure was observed with trivalent cations of Ga (CuS), In (CuS), Sc (FeAs, CrAs), V (FeAs), Cr (CuS, FeAs), Mn (CuS), and Fe (CuS) while the 32522 structure has been found for Sc (CuS, FeP, FeAs) and Fe (CuS, CuSe, AgS, and AgSe) [3, 4, 14, 18, 20 – 25]. In cases where both structures are formed, the differences in the *a* cell parameters are not so great as to suggest that the *anti*-fluorite layers should be strained too much to contribute to one structure but not to the other. Evidently, more thorough investigations of both structure types are necessary to discover the trends in relative stability of the structures in question. However, $Ca^{2+}$ (and probably $Eu^{2+}$) can be employed **more widely** to partially substitute for $Sr^{2+}$ in the structures of existing FeAs superconductors as a "chemical press", to increase the transition temperature.

### 5. Acknowledgements

The authors would like to thank Prof. V.M. Pudalov and Prof. E.V. Antipov for helpful discussion. This work was partially supported by the Ministry of Science and Education of Russian Federation under the State contracts P-279 (D.O.C. and S.M.K.), P2306 and 02.552.11.7093 (A.V.S. and O.E.O). A.V.S. would like to thank LPI Educational Research Center for the support. The support of the Russian Foundation for Basic Research is acknowledged (Grants No 10-03-00681-a, 09-02-01370).

**References**

[1] Y. Kamihara, T. Watanabe, M. Hirano, H. Hosono, J. Am. Chem. Soc. 130, 3296 (2008).

[2] S.J. Clarke, P. Adamson, S.J.C. Herkelrath, O.J. Rutt, D.R. Parker, M.J. Pitcher, C.F. Smura, Inorg. Chem. 47, 8473 (2008).

[3] X. Zhu, F. Han, G. Mu, P. Cheng, B. Shen, B. Zeng, H.-H. Wen, Phys. Rev. B 79, 220512 (R) (2009).

[4] X. Zhu, F. Han, G. Mu, B. Zeng, P. Cheng, B. Shen, H.H. Wen, Phys. Rev. B 79, 024516 (2009).

[5] V. Johnson and W. Jeitschko, J. Solid State Chem. 11, 161 (1974).

[6] D.O. Charkin and X.N. Zolotova, Cryst. Rev. 13, 201 (2007).

[7] E. Parthé, S. Hu, J. Solid State Chem. 174, 165 (2003).

[8] C.S. Knee, M.A.L. Field, M.T. Weller, Solid State Sci. 6, 443 (2004).

[9] Z. Hiroi, N. Kobayashi, M. Takano, Physica C 266, 191 (1996).

[10] T. Sowa, M. Hiratani, K. Miyauchi, J. Solid State Chem. 84, 178 (1990).

[11] Y. Shimizu, H. Ogino, N. Kawaguchi, K. Kishio, J. Shimoyama, arXiv:1006.3769 (unpublished).

[12] H. Ogino, S. Sato, K. Kishio, J. Shimoyama, T. Tohei, Y. Ikuhara, Appl. Phys. Expr. 3 (2010) 063103.

[13] H. Ogino, Y. Shimizu, K. Ushiyama, N. Kawaguchi, K. Kishio, J. Shimoyama, arXiv:1006.2355 (unpublished).

[14] W.J. Zhu, P.H. Hor, Inorg. Chem. 36, 3576 (1997).

[15] W.J. Zhu, P.H. Hor, J. Solid State Chem. 134, 128 (1997).

[16] TOPAS, version 3; Bruker AXS: Karlsruhe, Germany, (2005).

[17] J.M. Delgado, G. Diaz de Delgado, M. Quintero, J.C. Woolley, Mat. Res. Bull. 27, 367 (1992).

[18] W.J. Zhu, P.H. Hor, J. Solid State Chem. 153, 26 (2000).

[19] P. Quebe, L.J. Terbüchte, W. Jeitschko, J. Alloys Compd. 302, 70 (2000).

[20] L. Cario, A. Lafond, T. Morvan, H. Kabbour, G. André, P. Palvadeau, Solid State Sci. 7, 936 (2005).

[21] K. Otzschi, H. Ogino, J.-I. Shimoyama, K. Kishio, J. Low Temp. Phys. 117, 729 (1999).

[22] Y. L. Xie, R. H. Liu, T. Wu, G. Wu, Y. A. Song, D. Tan, X. F. Wang, H. Chen, J. J. Ying, Y. J. Yan, Q. J. Li, X. H. Chen, EPL 86, 57007 (2009).

[23] M. Tegel, F. Hummel, S. Lackner, I. Schellenberg, R. Poettgen, D. Johrendt, Z. Anorg. Allg. Chem. 635, 2242 (2009).

[24] H. Ogino, Y. Matsumura, Y. Katsura, K. Ushiyama, S. Horii, K. Kishio, J. Shimoyama, Supercond. Sci. Technol. 22, 075008 (2009).

[25] H. Ogino, Y. Katsura, S. Horii, K. Kishio, J. Shimoyama, Supercond. Sci. Technol. 22, 085001 (2009).

10Tables

**Table 1.** Crystallographic data for $Ca_2CuFeO_3S$ and $Ca_2CuFeO_3Se$

| Compound | $Ca_2CuFeO_3S$ | $Ca_2CuFeO_3Se$ |
|---|---|---|
| Formula weight | 559.225 | 653.012 |
| Crystal system | Tetragonal | |
| Space group | P4/nmm (#129) | |
| Cell parameters | | |
| $a$, Å | 3.8269(1) | 3.8596(1) |
| $c$, Å | 14.9482(3) | 15.2985(2) |
| V, Å$^3$ | 218.91(1) | 227.89(1) |
| Calculated density (g/cm$^3$) | 4.24 | 4.76 |
| Diffractometer | STOE | |
| Radiation | $CuK\alpha_1$ ($\lambda = 1.54056$Å) | |
| $2\vartheta$ range | 5-110 | 5-110 |
| No. of reflexions | 116 | 121 |
| No. of data points | 10499 | 10499 |
| No. of structural parameters | 13 | 17 |
| No. of overall parameters | 37 | 46 |
| Analyzing package | Topas [13] | |
| Reliability factors: | | |
| $R_p$ | 0.014 | 0.011 |
| $R_{wp}$ | 0.015 | 0.014 |
| $\chi^2$ | 1.33 | 1.37 |

**Table 2.** Refined atomic coordinates for the $Ca_2CuFeO_3Ch$ compounds

| Compound | $Ca_2CuFeO_3S$ | | $Ca_2CuFeO_3Se$ | |
|---|---|---|---|---|
| Atom | $z$ | B, Å$^2$ | $z$ | B, Å$^2$ |
| Ca1 (0.75, 0.75, $z$) | 0.1890(3) | 0.52(8) | 0.1981(1) | 0.66(4) |
| Ca2 (0.75, 0.75, $z$) | 0.4130(2) | 0.52(9) | 0.4163(1) | 1.28(5) |
| Fe (0.25, 0.25, $z$) | 0.3099(2) | 0.51(6) | 0.3167(1) | 0.71(4) |
| Cu (0.25, 0.75, 0) | 0 | 1.53(5) | 0 | 1.48(3) |
| Ch (0.25, 0.25, $z$) | 0.0981(3) | 0.5(1) | 0.1044(1) | 0.92(3) |
| O1 (0.25, 0.75, $z$) | 0.2889(3) | 1.0[a] | 0.2961(2) | 1.10(7)[b] |
| O2 (0.25, 0.25, $z$) | 0.4392(6) | 1.0[a] | 0.4406(3) | 1.10(7)[b] |

a - were fixed; b - were constrained to be equal.

**Table 3.** Selected bond lengths for $Ca_2FeO_3Cl$, $Ca_2CuFeO_3S$, $Sr_2CuFeO_3S$, and $Ca_2CuFeO_3Se$

| Compound | $Ca_2FeO_3Cl$ [7] | $Ca_2CuFeO_3S$ | $Sr_2CuFeO_3S$ [15] | $Ca_2CuFeO_3Se$ |
|---|---|---|---|---|
| Ae1 – O1 x4 | 2.293 | 2.428(3) | 2.512(7) | 2.444(2) |
| Ae2 – O1 x4 | 2.397 | 2.664(3) | 2.758(8) | 2.666(3) |
| Ae2 – O2 x4 | 2.960 | 2.734(1) | 2.774(2) | 2.7542(7) |
| Fe – O1 x4 | 1.959 | 1.9389(7) | 1.984(2) | 1.9553(4) |
| Fe – O2 x1 | 1.854 | 1.933(9) | 1.91(1) | 1.895(5) |
| Cu – Ch x4 |  | 2.411(2) | 2.456(6) | 2.5055(6) |

**Figure Captions:**

**Figure 1.** The crystal structures (left to right) of (a) $Sr_3Fe_2O_5Cl_2$ and $Sr_3Fe_2Cu_2O_5S_2$; (b) $Sr_2FeO_3Cl$ and $Sr_2CuFeO_3S$ illustrating the "filling" concept.

**Figure 2.** (a) Observed (blue), calculated (red) and difference (gray) Rietveld plots for $Ca_2CuFeO_3S$. The vertical bars indicate the reflection positions. (b) Observed (blue), calculated (red) and difference (gray) Rietveld plots for $Ca_2CuFeO_3Se$. The vertical bars indicate the reflection positions. $CuFeSe_2$ was included in the refinement as impurity phase (content = 3.2%).

**Figure 3.** Temperature dependence of ac-susceptibility of (a) $Ca_2CuFeO_3S$ and (b) $Ca_2CuFeO_3Se$.

**Figure 4.** (a) Temperature dependence of resistivity of $Ca_2CuFeO_3S$. The inset shows a linear section of $R(T)$ dependence in $\ln r - 1/T$ coordinates typical for temperature activation transport mechanism. (b) Temperature dependence of resistivity of $Ca_2CuFeO_3Se$. The insert shows a linear section of $R(T)$ dependence in $\ln r - 1/T^4$ coordinates typical for variable-range hopping transport.

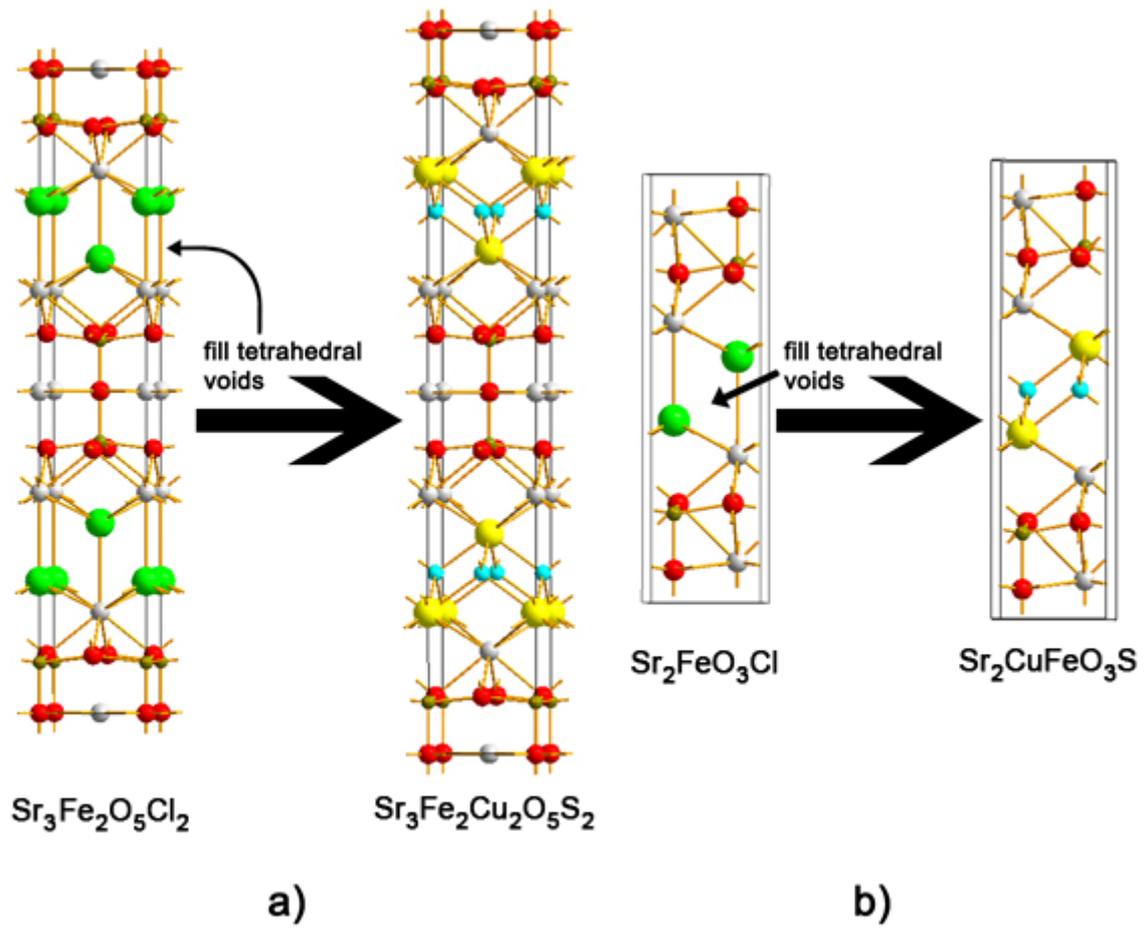

Figure 1

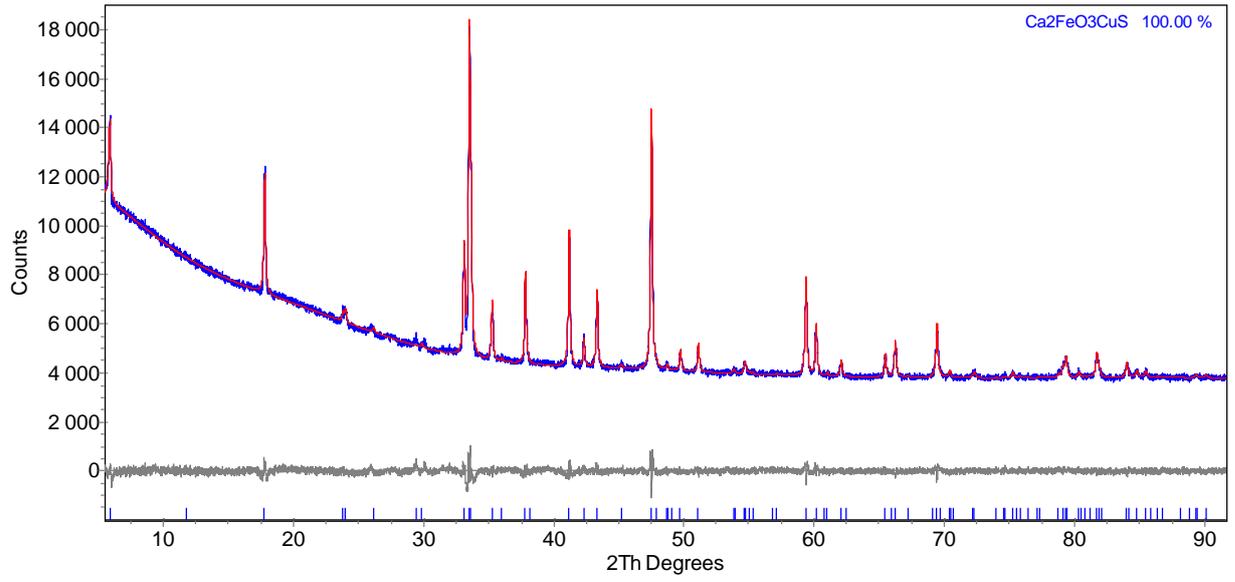

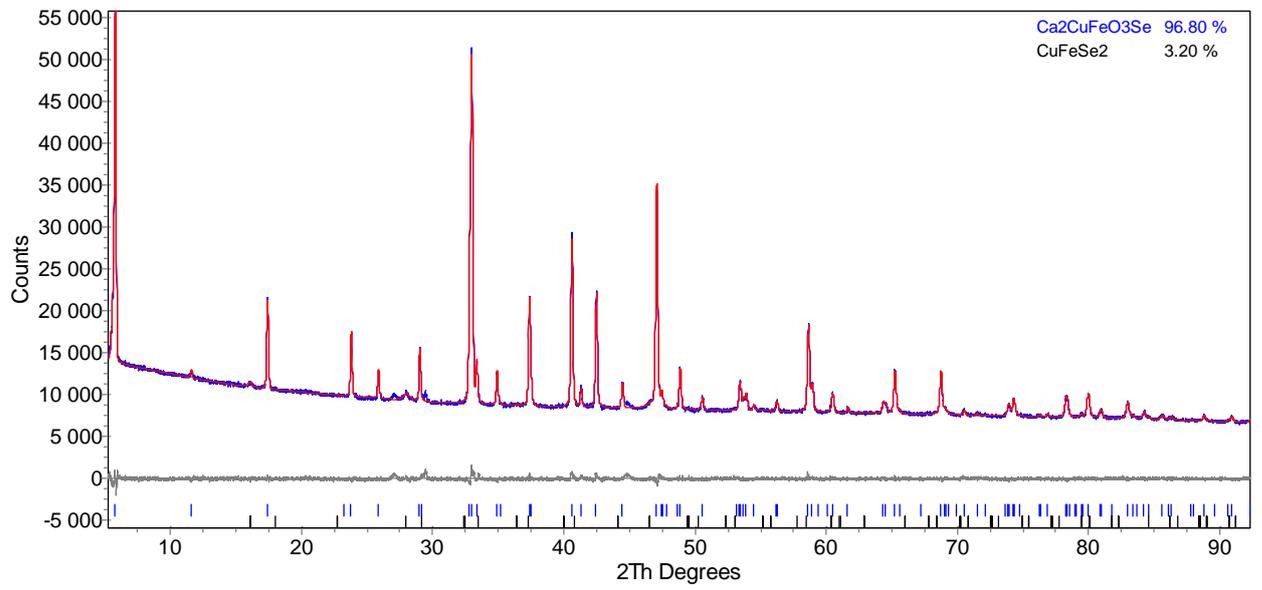

Fig. 2.

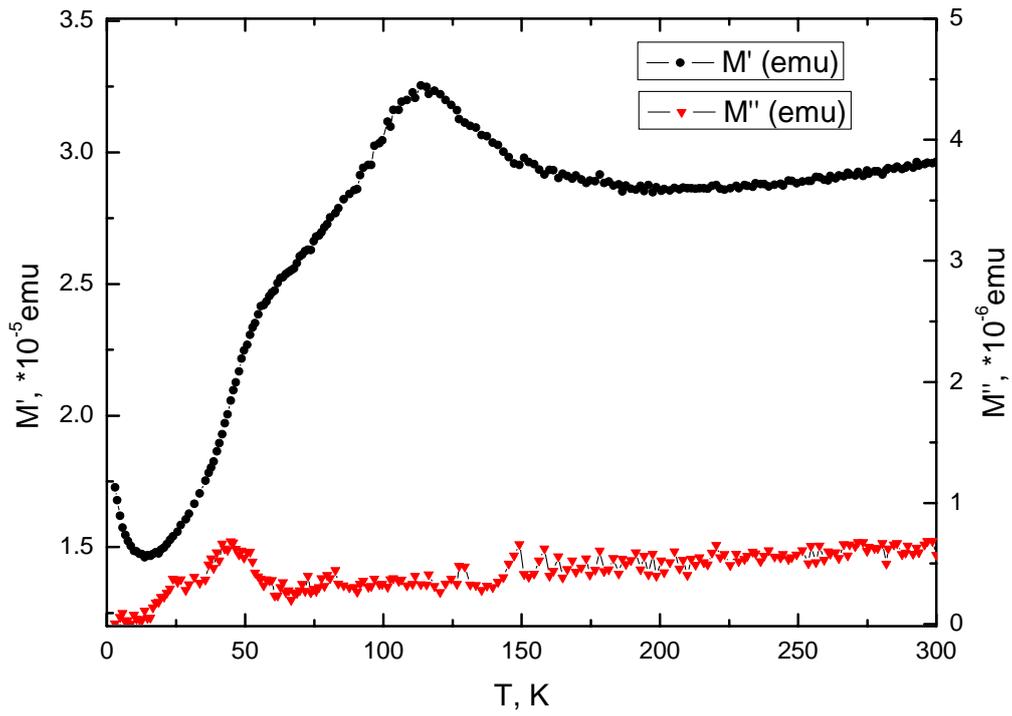

a)

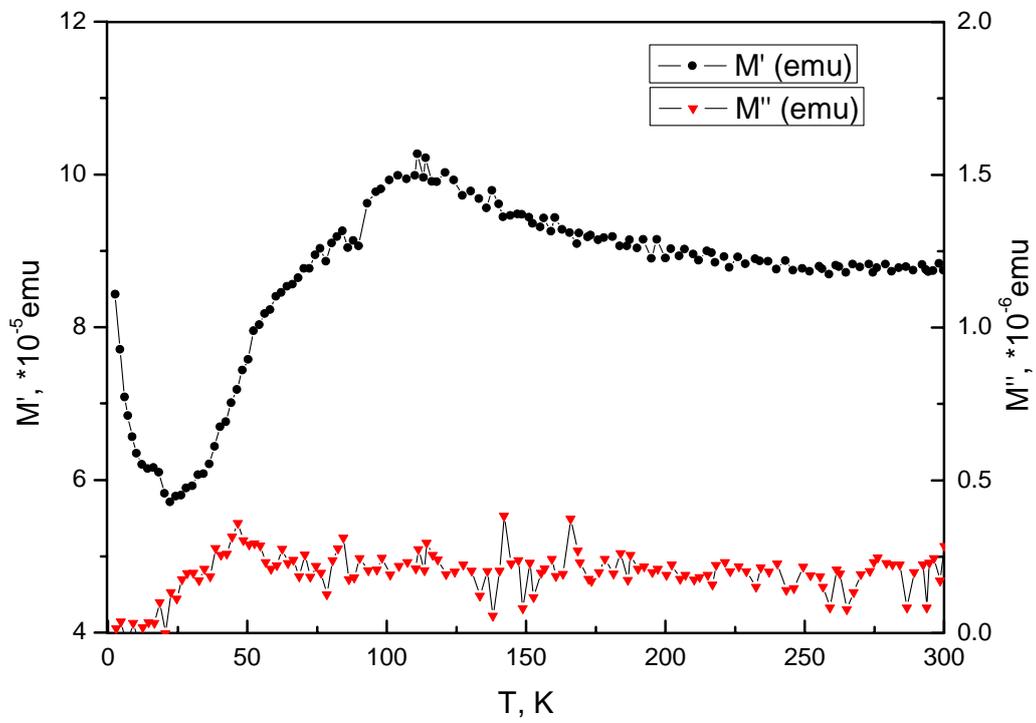

b)

Fig. 3.

``

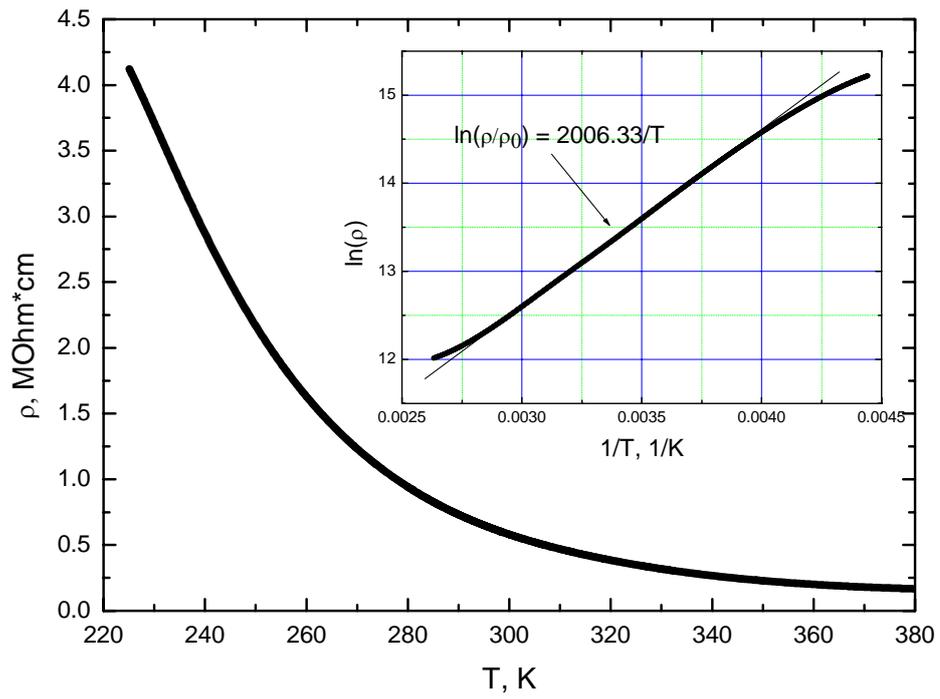

a)

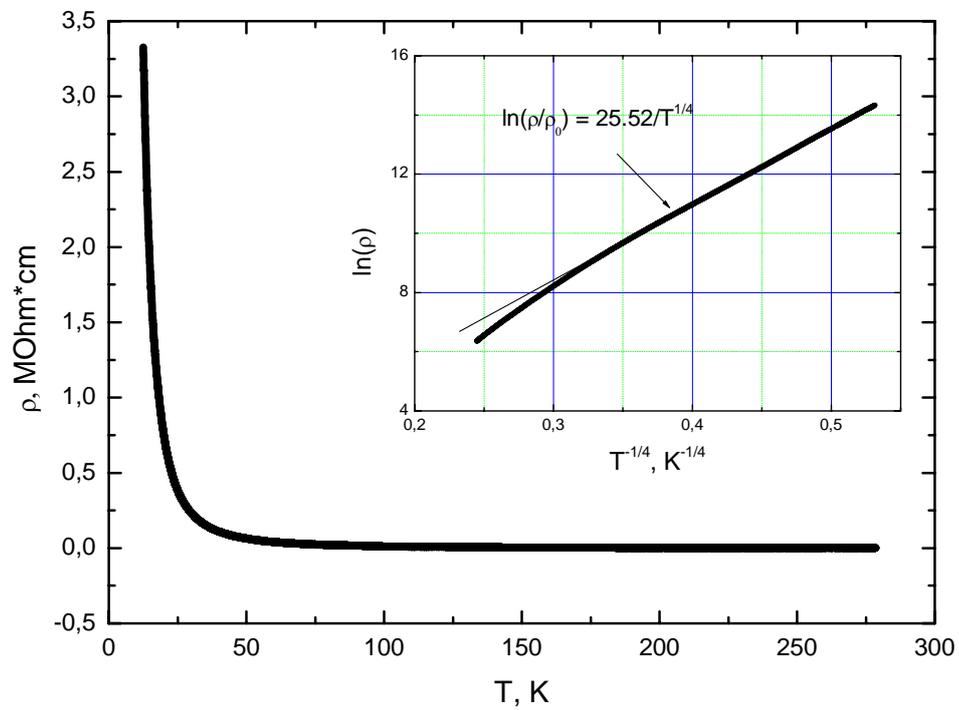

b)

Fig. 4.